\documentstyle[preprint,revtex,eqsecnum]{aps}
\begin{document}
\draft
\baselineskip=1.1\baselineskip
\begin{title}
Five--loop $\sqrt\epsilon$--expansions for random Ising model \\
and marginal spin dimensionality for cubic systems.
\end{title}
\author{B.~N.~Shalaev$^1$, ~S.~A.~Antonenko$^2$, 
~A.~I.~Sokolov$^2$}
\begin{instit}
$^1$A.~F.~Ioffe Physical and Technical Institute, Russian Academy
of Sciences, \\
Politekhnicheskaya str. 26, Saint Petersburg 194021, RUSSIA \\
$^2$Saint Petersburg Electrotechnical University, Professor 
Popov str. 5, \\ Saint Petersburg 197376, RUSSIA
\end{instit}

\begin{abstract}
The $\sqrt\epsilon$--expansions for critical exponents of the
weakly--disordered Ising model are calculated up to the five-loop
order and found to possess coefficients with irregular signs and 
values.  The estimate $n_c = 2.855$ for the marginal spin 
dimensionality of the cubic model is obtained by the Pade--Borel 
resummation of corresponding five--loop $\epsilon$--expansion.

~Cubic systems; ~Marginal spin 
dimensionality

\end{abstract}

\newpage

\section{ INTRODUCTION }
\label{sec:1}

The critical behaviour of weakly--disordered quenched systems
undergoing continuous phase transitions is of great interest.
The study of critical properties of random spin systems in which
the local energy density couples to quenched disorder has a long
history going back to the classical papers by A. B. Harris and
T. C. Lubensky \cite{HL,TCL} and D. E. Khmelnitskii
\cite{KDE}. They initiated a considerable progress in studying
disordered systems by applying the conventional
field--theoretical renormalization--group (RG) approach based on
the standard scalar $\varphi^4$ theory with $n$-component order
parameter in $(4 - \epsilon)$ space dimensions.
At present it is commonly believed that the RG approach provides
a thorough understanding how weak disorder affects thermodynamic
properties of random systems in a close vicinity of the
Curie point.

According to the Harris criterion, critical exponents of
weakly disordered Ising model should differ from those for the
pure system \cite{ABH}. The first regular method for calculating
their values, famous $\sqrt{\epsilon}$--expansion, was invented
more than 20 years ago \cite{HL,TCL,KDE} but
turned out to be numerically ineffective, at least in lower
orders in $\sqrt{\epsilon}$. Fair numerical estimates for
critical exponents of the random Ising model (RIM) were obtained
in the framework of the renormalization--group approach in
three dimensions from two--loop \cite{GJ}, three--loop \cite{MS}
and four--loop \cite{MSS,IOM} RG expansions.
The RG method at fixed dimensions proved also to be efficient 
when used to calculate critical exponents of the two--dimensional 
RIM and the marginal value $n_c$ of the order parameter
dimensionality $n$ for the cubic model \cite{IOM,MSI,MSF,BNS2};
as is known, $n_c$ separates the region where the system becomes
effectively isotropic approaching the critical point ($n < n_c$)
from that of the essentially anisotropic critical behaviour
($n > n_c$).

Recently, H. Kleinert and V. Schulte--Frohlinde have found the 
RG functions for the $(4 - \epsilon)$--dimensional hypercubic 
model in the five--loop approximation \cite{HK}. To obtain the 
RG series of unprecedented length for the model with two quartic
coupling constants these authors have employed the early results
of five--loop RG calculations for $O(n)$--symmetric $\varphi^4$
field theory \cite{DIK}. It is well known that in the replica
limit $n \rightarrow 0$ the scalar $\varphi^4$ field theory with
$O(n)$--symmetric and hypercubic self--interactions describes
the critical behaviour of the RIM provided the coupling
constants have proper signs \cite{KZ}. Hence, the RG expansions 
obtained in \cite{HK} may be used for calculation of RIM critical
exponents as power series in $\sqrt{\epsilon}$ or,
more precisely, for extension of known three--loop expansions
\cite{BNS1,JK} up to five--loop
($\epsilon^2$) terms. To find such five--loop expansions
is the main goal of this paper.

Another goal is to get numerical estimate for $n_c$ starting
from the $\epsilon$--expansion for this quantity obtained
in Ref. \cite{HK} and to compare this estimate with its analogs
found earlier from RG expansions in three dimensions 
\cite{MSS,MSI,MSF}. If the results given by these two
approaches are in accord we shall be able, at last, to answer
the old question: is a cubic crystal effectively ``isotropic''
at the critical point?

\section{RG FUNCTIONS AND CRITICAL EXPONENTS
FOR RANDOM ISING MODEL}
\label{sec:2}

We begin with the standard Landau Hamiltonian for a model with
hypercubic anisotropy, describing numerous magnetic and
structural phase transitions in solids. It reads:
\begin{eqnarray}
H &=& \int d^d x \biggl[ \frac{1}{2} (\partial_{\mu} \varphi)^2 
+ \frac{1}{2} m_o^2 \varphi^2 + \frac{1}{4 !} u_o (\varphi^2)^2 
+ \frac{1}{4 !} v_{o} \sum_{a=1}^n \varphi_a^4 \biggr] \ \ , 
\nonumber \\ 
\varphi^2 &=& \sum_{a=1}^n \varphi_a^2 \ \ , \qquad 
(\partial_{\mu} \varphi)^2 = \sum_{a=1}^n (\partial_{\mu} 
\varphi_{a})^2 \ \ , 
\label{a}
\end{eqnarray}
\noindent
where $\varphi$ is an $n$--component order parameter, 
$m_o^2 \sim \tau$, with $\tau = \frac{T - T_c}{T_c}$ being 
the reduced
deviation from the mean--field transition temperature.
In the replica limit, the Hamiltonian Eq.~(\ref{a}) is known to
describe the RIM provided $v_o > 0$ and $u_o < 0$.

If we set $n$ to zero in formulas obtained in \cite{HK},
we arrive, directly, at five--loop $\beta$--functions and critical
exponents for the model under consideration:
\begin{eqnarray}
\frac{\beta_u (u,v)}{u} &=& - \epsilon + \frac{8}{3} u + 2v
- \frac{14}{3} u^2 - \frac{22}{3} uv - \frac{5}{3} v^2 
+ u^3 \biggl( \frac{370}{27} + \frac{88}{9} \zeta(3) \biggr) 
+ u^2 v \biggl( \frac{659}{18} + \frac{64}{3} \zeta(3) \biggr) 
\nonumber \\ 
&+& uv^2 \biggl( \frac{107}{4} + 8 \zeta(3) \biggr) + 7 v^3 
+ u^4 \biggl( -\frac{24581}{486} - \frac{4664}{81} \zeta(3) 
+ \frac{352}{27} \zeta(4) - \frac{2480}{27} \zeta(5) \biggr) 
\nonumber \\ 
&+& u^3 v \biggl( -\frac{15967}{81} - \frac{4856}{27} \zeta(3) 
+ \frac{340}{9} \zeta(4) - \frac{2560}{9} \zeta(5) \biggr) 
+ u^2 v^2 \biggl( -\frac{13433}{54} - \frac{1456}{9} \zeta(3) 
\nonumber \\ 
&+& \frac{64}{3} \zeta(4) - \frac{2000}{9} \zeta(5) \biggr) 
+ uv^3 \biggl( - \frac{4867}{36} - 50 \zeta(3)
- 8 \zeta(4) - \frac{160}{3} \zeta(5) \biggr) 
\nonumber \\
&+& v^4 \biggl( - \frac{477}{16} - 3 \zeta(3) - 6 \zeta(4) 
\biggr) + u^5 \biggl( \frac{17158}{81} + \frac{27382}{81} \zeta(3) 
+ \frac{1088}{27} \zeta(3)^2 - \frac{880}{9} \zeta(4) 
\nonumber \\ 
&+& \frac{55028}{81} \zeta(5) - \frac{6200}{27} \zeta(6) 
+ \frac{25774}{27} \zeta(7) \biggr) 
+ u^4 v \biggl( \frac{537437}{486} + \frac{116759}{81} \zeta(3) 
+ \frac{3148}{27} \zeta(3)^2 
\nonumber \\ 
&-& \frac{10177}{27} \zeta(4) + \frac{75236}{27} \zeta(5) 
- \frac{24050}{27} \zeta(6) + \frac{11564}{3} \zeta(7) \biggr) 
+ u^3 v^2 \biggl( \frac{1314497}{648} 
\nonumber \\ 
&+& \frac{171533}{81} \zeta(3) + \frac{1384}{27} \zeta(3)^2 
- \frac{23105}{54} \zeta(4) 
+ \frac{96794}{27} \zeta(5)- \frac{25400}{27} \zeta(6) 
+ \frac{14210}{3} \zeta(7) \biggr) 
\nonumber \\ 
&+& u^2 v^3 \biggl( \frac{2281727}{1296} 
+ \frac{37789}{27} \zeta(3) - \frac{544}{9} \zeta(3)^2 
- \frac{337}{3} \zeta(4) + \frac{17444}{9} \zeta(5) 
- \frac{1600}{9} \zeta(6) 
\nonumber \\ 
&+& 2352 \zeta(7) \biggr) + u v^4 \biggl( \frac{1336801}{1728} 
+ \frac{5495}{12} \zeta(3) - \frac{190}{3} \zeta(3)^2 
+ \frac{141}{2} \zeta(4) 
\nonumber \\
&+& \frac{1145}{3} \zeta(5) + \frac{575}{3} \zeta(6) 
+ 441 \zeta(7) \biggr) + v^5 \biggl( \frac{158849}{1152} 
+ \frac{1519}{24} \zeta(3) 
\nonumber \\
&-& 18 \zeta(3)^2 + \frac{65}{2} \zeta(4) + 2 \zeta(5)
+ 75 \zeta(6) \biggr) \ \ , 
\label{b}
\end{eqnarray}

\begin{eqnarray}
\frac{\beta_v (u,v)}{v} &=& - \epsilon + 3v + 4u 
- \frac{17}{3} v^2 - \frac{46}{3} uv - \frac{82}{9} u^2 
+ v^3 \biggl( \frac{145}{8} + 12 \zeta(3) \biggr) 
+ uv^2 \biggl( \frac{131}{2} + 48 \zeta(3) \biggr) 
\nonumber \\ 
&+& u^2 v \biggl( \frac{325}{4} + 64 \zeta(3) \biggr)
+ u^3 \biggl( \frac{821}{27} + \frac{224}{9} \zeta(3) \biggr) 
+ v^4 \biggl( - \frac{3499}{48} - 78 \zeta(3) +18 \zeta(4) 
\nonumber \\ 
&-& 120 \zeta(5) \biggr) + uv^3 \biggl( - \frac{1004}{3} 
- 387 \zeta(3) + 96 \zeta(4) - 600 \zeta(5) \biggr) 
+ u^2 v^2 \biggl( - \frac{10661}{18} 
\nonumber \\
&-& 724 \zeta(3) + 184 \zeta(4) - \frac{3440}{3} \zeta(5) 
\biggr) + u^3 v \biggl( - \frac{12349}{27} 
- \frac{5312}{9} \zeta(3) 
\nonumber \\ 
&+& \frac{440}{3} \zeta(4) -960 \zeta(5) \biggr) 
+ u^4 \biggl( - \frac{19679}{162} - 168 \zeta(3) 
+ 40 \zeta(4)- \frac{7280}{27} \zeta(5) \biggr) 
\nonumber \\ 
&+& v^5 \biggl( \frac{764621}{2304} + \frac{7965}{16} \zeta(3) 
+ 45 \zeta(3)^2 - \frac{1189}{8} \zeta(4) + 987 \zeta(5) 
\nonumber \\
&-& \frac{675}{2} \zeta(6) + 1323 \zeta(7) \biggr) 
+ uv^4 \biggl( \frac{1067507}{576} 
+ \frac{35083}{12} \zeta(3) + 288 \zeta(3)^2 
\nonumber \\ 
&-& \frac{3697}{4} \zeta(4) + 5920 \zeta(5) 
- 2100 \zeta(6) + 7938 \zeta(7) \biggr) 
\nonumber \\ 
&+& u^2 v^3 \biggl( \frac{3633377}{864} 
+ \frac{125459}{18} \zeta(3) + \frac{2266}{3} \zeta(3)^2 
- 2263 \zeta(4) 
\nonumber \\
&+& 14328 \zeta(5)- \frac{15575}{3} \zeta(6)+19404 \zeta(7) 
\biggr) + u^3 v^2 \biggl( \frac{9309907}{1944} 
+ \frac{224804}{27} \zeta(3) \nonumber \\ 
&+& \frac{3032}{3} \zeta(3)^2 
- \frac{73018}{27} \zeta(4)+ \frac{155692}{9} \zeta(5) 
-6300 \zeta(6) + 23912 \zeta(7) \biggr) 
\nonumber \\ 
&+& u^4 v \biggl( \frac{1279979}{486}+ \frac{784621}{162} \zeta(3) 
+ \frac{18154}{27} \zeta(3)^2 - \frac{83837}{54} \zeta(4) 
+ \frac{275510}{27} \zeta(5) 
\nonumber \\ 
&-& \frac{98975}{27} \zeta(6) + \frac{43120}{3} \zeta(7) \biggr) 
+ u^5 \biggl( \frac{389095}{729} + \frac{259358}{243} \zeta(3) 
+ \frac{13288}{81} \zeta(3)^2 
\nonumber \\ 
&-& \frac{27166}{81} \zeta(4)+ \frac{179696}{81} \zeta(5) 
- \frac{63500}{81} \zeta(6) + \frac{28420}{9} \zeta(7) \biggr) \ \ , 
\label{c}
\end{eqnarray}

\begin{eqnarray}
\eta(u,v) &=& \frac{1}{9} u^2 + \frac{1}{3} uv 
+ \frac{1}{6} v^2 - \frac{2}{27} u^3 
- \frac{1}{3} u^2 v - \frac{3}{8} uv^2 - \frac{1}{8} v^3 
+ u^4 \frac{125}{324} 
\nonumber \\
&+& u^3 v \frac{125}{54} + u^2 v^2 \frac{145}{36} 
+ \frac{65}{24} uv^3 + v^4 \frac{65}{96} + u^5 \biggl( 
- \frac{1204}{729} + \frac{46}{243} \zeta(3) 
- \frac{44}{81} \zeta(4) \biggr) 
\nonumber \\ 
&+& u^4 v \biggl( - \frac{3010}{243} 
+ \frac{115}{81} \zeta(3) - \frac{110}{27} \zeta(4) \biggr) 
+ u^3 v^2 \biggl( - \frac{58177}{1944} + \frac{191}{54} \zeta(3) 
- \frac{260}{27} \zeta(4) \biggr) 
\nonumber \\ 
&+& u^2 v^3 \biggl( - \frac{13741}{432} + \frac{67}{18} \zeta(3) 
- 10 \zeta(4) \biggr) + uv^4 \biggl( - \frac{18545}{1152} 
+ \frac{15}{8} \zeta(3) - 5 \zeta(4) \biggr) 
\nonumber \\ 
&+& v^5 \biggl( - \frac{3709}{1152} + \frac{3}{8} \zeta(3) 
- \zeta(4) \biggr) \ \ , 
\label{d}
\end{eqnarray}

\begin{eqnarray}
\nu(u,v)^{-1} &=& 2 - \frac{2}{3} u - v + \frac{5}{9} u^2 
+ \frac{5}{3} uv + \frac{5}{6} v^2 - \frac{37}{18} u^3 
- \frac{37}{4} u^2 v - \frac{251}{24} uv^2 
\nonumber \\ 
&-& \frac{7}{2} v^3 - u^4 \biggl( - \frac{7765}{972} 
- \frac{68}{81} \zeta(3) - \frac{44}{27} \zeta(4) \biggr) 
- u^3 v \biggl( - \frac{7765}{162} 
\nonumber \\ 
&-& \frac{136}{27} \zeta(3) - \frac{88}{9} \zeta(4) \biggr) 
- u^2 v^2 \biggl( - \frac{9199}{108} - \frac{26}{3} \zeta(3) 
- \frac{52}{3} \zeta(4) \biggr) 
\nonumber \\ 
&-& uv^3 \biggl( - \frac{4243}{72} 
- \frac{19}{3} \zeta(3) - 12 \zeta(4) \biggr) 
- v^4 \biggl( - \frac{477}{32} 
- \frac{3}{2} \zeta(3) - 3 \zeta(4) \biggr) \ \ , 
\label{e}
\end{eqnarray}

\noindent
where numbers $\zeta(3) = 1.202056903$; $\zeta(4) = 1.082323234$;
$\zeta(5) = 1.036927755$; $\zeta(6) = 1.017343062$;
$\zeta(7)=1.008349277$ are the values of the Riemann
$\zeta$--function.
The  coordinates of the random infrared-free fixed point being
zeros of the above $\beta$--functions, they are found in the
following form \cite{BNS1,JK}:

\begin{equation}
u^* = - A \sqrt{\epsilon} + B \epsilon + C {\sqrt{\epsilon}}^3 
+ D \epsilon^2 + K {\sqrt{\epsilon}}^5 + \ldots \ \ , 
\label{f}
\end{equation}

\begin{equation}
v^* = \frac{4}{3} A \sqrt{\epsilon} + (F - \frac{4}{3} B) \epsilon 
+ (G - \frac{4}{3} C) {\sqrt{\epsilon}}^3 
+ (H - \frac{4}{3} D) \epsilon^2 
+ (L - \frac{4}{3} K) {\sqrt{\epsilon}}^5 + \ldots \ \ . 
\label{ff}
\end{equation}

\noindent
Substituting $u^*$ and $v^*$ into the beta--functions, we get 
algebraic equations which decouple into 4 pairs of equations
for A and F, B and G, C and H, and, at last, for D and L,
respectively. Note, that the coefficient K may be computed only
in the sixth--loop approximation. After straightforward but
cumbersome calculations one is led to the following expressions:

\begin{eqnarray}
u^* &=& - \sqrt{\epsilon} \frac{3 \sqrt{318}}{106} 
+ \epsilon \biggl( \frac{567 \zeta(3)}{2809} 
+ \frac{990}{2809} \biggr) 
- {\sqrt{\epsilon}}^3 {9 \sqrt{318} \over {252495392}} 
\biggl( 317520 \zeta(3)^2 + 775536 \zeta(3) 
\nonumber \\ 
&+& 5(114365-137376 \zeta(5)) \biggr) + \epsilon^2 
{9 \over {3345563944}} \biggl( 96018048 \zeta(3)^3 
+ 326732616 \zeta(3)^2 \nonumber \\ 
&+& 6 \zeta(3) (26149279 - 64910160 \zeta(5)) 
+ 29477646 \zeta(4) 
\nonumber \\ 
&-& 309476010 \zeta(5)-367914393 \zeta(7) 
+101727760 \biggr) \ \ ,
\label{g}
\end{eqnarray}

\begin{eqnarray}
v^{*} &=& \sqrt{\epsilon} \frac{2 \sqrt{318}}{53} 
- 36 \epsilon \frac{(21 \zeta(3)+19)}{2809} 
+ {\sqrt{\epsilon}}^3 \biggl( 
\frac{119070 \sqrt{318} \zeta(3)^2}{7890481}
+ \frac{261252 \sqrt{318} \zeta(3)}{7890481} 
\nonumber \\ 
&-& \frac{4860 \sqrt{318} \zeta(5)}{148877} 
+ \frac{3 \sqrt{83789895038142}}{63123848} \biggr) 
- \epsilon^2 {27 \over {836390986}}
\biggl( 10668672 \zeta(3)^3 
\nonumber \\ 
&+& 33819408 \zeta(3)^2 + 2 \zeta(3)(5195389 
- 21636720 \zeta(5)) + 3870802 \zeta(4) 
\nonumber \\ 
&-& 30191450 \zeta(5) - 40879377 \zeta(7) + 2892644 \biggr) \ \ . 
\label{h}
\end{eqnarray}

\noindent
The impure fixed point has been proved to be stable in the
framework of the perturbation theory in $\sqrt{\epsilon}$.
Hence, we have to insert $u^*$ and $v^*$ into the critical
exponents series.
The expansions for critical exponents in powers of $\sqrt{\epsilon}$
being the eventual goal of this study are as follows:

\begin{eqnarray}
\eta &=& - \frac{\epsilon}{106} + {\sqrt{\epsilon}}^3
{9 \sqrt{318} \over {148877}} \biggl( 7 \zeta(3) + 24 \biggr) 
- \epsilon^2 {27 \over {63123848}} \biggl( 21168 \zeta(3)^2 
\nonumber \\ 
&+& 76040 \zeta(3) - 38160 \zeta(5) + 22469 \biggr) 
+ {\sqrt{\epsilon}}^5 \biggl( \frac{15752961 \sqrt{318} \zeta(3)^{3}}
{22164361129} + \frac{70314804 \sqrt{318} \zeta(3)^2}
{22164361129} \nonumber \\ 
&+& 27 \sqrt{318} \zeta(3) \frac{26313923 
- 33657120 \zeta(5)}{354629778064} 
+ \frac{189 \sqrt{318} \zeta(4)}{595508} 
\nonumber \\ 
&-& \frac{4725 \sqrt{467895342} \zeta(5)}{1672781972} 
- \frac{130977 \sqrt{318} \zeta(7)}{63123848} 
+ \frac{2997 \sqrt{152112323838}}{44328722258} \biggr) \ \ , 
\label{k}
\end{eqnarray}

\begin{eqnarray}
\nu &=& \frac{1}{2} + \sqrt{\epsilon} \frac{\sqrt{318}}{212} 
+ \epsilon \frac{535 - 756 \zeta(3)}{22472} 
+ {\sqrt{\epsilon}}^3 \biggl( \frac{59535 \sqrt{318} \zeta(3)^2}
{31561924} + \frac{39555 \sqrt{318} \zeta(3)}{15780962} \nonumber \\ 
&-& \frac{1215 \sqrt{318} \zeta(5)}{297754} 
+ \frac{397 \sqrt{705044478}}{504990784} \biggr) 
- \epsilon^{2} {1 \over {6691127888}} 
\biggl( 288054144 \zeta(3)^{3}+679447440 \zeta(3)^{2} 
\nonumber \\ 
&-& 45 \zeta(3) (25964064 \zeta(5) + 8113195) + 168826518 \zeta(4) 
\nonumber \\ 
&-& 401571990 \zeta(5) - 7 (157677597 \zeta(7) + 9164941) \biggr) 
\ \ . 
\label{l}
\end{eqnarray}

\noindent
The $\sqrt{\epsilon}$ and $\epsilon$ terms in 
Eq.~(\ref{k}) and first three terms 
in Eq.~(\ref{l}) coincide with those calculated earlier
\cite{BNS1,JK} while the rest are essentially new.
Estimating coefficients in the above expansions numerically we
obtain:

\begin{equation}
\eta = - 0.0094339622 \epsilon + 0.034943501 {\sqrt{\epsilon}}^3 
- 0.044864982 \epsilon^2 + 0.021573216 {\sqrt{\epsilon}}^5 \ \ , 
\label{m}
\end{equation}

\begin{equation}
\nu = \frac{1}{2} + 0.084115823 \sqrt{\epsilon} 
- 0.016632032 \epsilon + 0.047753505 {\sqrt{\epsilon}}^3 
+ 0.27258431 \epsilon^2 \ \ . 
\label{n}
\end{equation}

Corresponding $\sqrt{\epsilon}$--expansion for the susceptibility
exponent reads: 

\begin{equation}
\gamma = 1 + 0.16823164 \sqrt{\epsilon} - 0.028547082 \epsilon 
+ 0.078828812 {\sqrt{\epsilon}}^3 + 0.56450485 \epsilon^2 \ \ . 
\label{nn}
\end{equation}

The most striking feature of the series just obtained is the
quite irregular behaviour of their coefficients. It may be
considered as one of the main reasons of failure of our attempts
to apply various resummation techniques to these expansions.
Indeed, the numerical results found by means of several methods
based on the Borel transformation turned out both to contradict
to exact inequalities and to differ markedly from estimates
given by 3D RG analysis \cite{MS,MSS,IOM} 
and by computer simulations \cite{HF}. For example, numerical
estimates for the exponent $\nu$ thus obtained lead, via
scaling relation $\alpha = 2 - D\nu$, to positive (and big!) 
values of the exponent $\alpha$, in obvious contradiction with 
the inequality $\alpha < 0$ proven for impure systems 
\cite{MA}. 

The ``odd'' behaviour of coefficients of 
the $\sqrt\epsilon$--expansions for critical exponents inherent 
in the RIM appears to be the rule rather than the exception,
this apparently indicates on the Borel non--summability 
of perturbative series. The point is that the analysis
of the perturbative expansions for the free energy of the
zero--dimensional (``toy'') model with quenched disorder has lead
to the conjecture that for disordered systems perturbative series
are Borel non--summable \cite{BCM}. Such an non--summability has 
then been related to Griffiths singularities \cite{MK}.
So, the irregularity of signs and values of coefficients of the
$\sqrt\epsilon$--expansions found may be regarded as a
manifestation of the Borel non--summability of these series.

\section{IS A CUBIC CRYSTAL ``ISOTROPIC'' AT THE CRITICAL POINT?}
\label{sec:3}

Much better situation takes place in the case of a pure system
with cubic anisotropy. Five--loop $\epsilon$--expansion for the
marginal spin dimensionality found by Kleinert and
Schulte--Frohlinde \cite{HK}

\begin{eqnarray}
n_c &=& 4 - 2\epsilon + {\epsilon}^2 \biggl(- \frac{5}{12} 
+ \frac{5\zeta(3)}{2} \biggr)
+ {\epsilon}^3 \biggl(- \frac{1}{72} + \frac{5\zeta(3)}{8} 
+ \frac{15\zeta(4)}{8} - \frac{25\zeta(5)}{3} \biggr) 
\nonumber \\
&+& {\epsilon}^4 \biggl(- \frac{1}{384} + 
\frac{93\zeta(3)}{128} - \frac{229\zeta^2(3)}{144} 
+ \frac{15\zeta(4)}{32} - \frac{3155\zeta(5)}{1728} 
- \frac{125\zeta(6)}{12} + \frac{11515\zeta(7)}{384} \biggr) 
\nonumber \\
&=& 4 - 2\epsilon + 2.58847559 {\epsilon}^2 
- 5.87431189 {\epsilon}^3 + 16.8270390 {\epsilon}^4
\label{o}
\end{eqnarray}
is seen to be alternating. Moreover, coefficients modulo of its
Borel transform are easily shown to monotonically decrease what
may be thought of as a manifestation of the Borel summability
of the original series.

Let us calculate $n_c$ for the three--dimensional system
applying the Pade--Borel resummation technique to the expansion
Eq.~(\ref{o}) and then putting $\epsilon$ equal to unity. In so
doing, however, it is very important to trace how sensitive is
the numerical estimate for $n_c$ thus obtained to the
perturbative order employed. That is why we calculate here $n_c$
not only in the five--loop approximation but also in three-- and
four--loop orders in $\epsilon$. Correspondingly, Pade
approximants [1/1], [2/1] and [3/1] are used for analytical
continuation of the Borel transforms of the original series.
The results obtained in three subsequent approximations mentioned
are as follows:

\begin{equation}
n_c^{(3)} = 3.004, \qquad \quad
n_c^{(4)} = 2.918, \qquad \quad
n_c^{(5)} = 2.855.
\label{p}
\end{equation}

These estimates are seen to behave quite regularly. They decrease
with increasing the order of the approximation demonstrating the
tendency to go deeper and deeper below 3. This fact is in
agreement with the conclusion about the character of critical
behaviour of cubic crystals made earlier on the base of
higher--order RG calculations in three dimensions \cite{MSS,MSI,MSF}. 
Indeed, two--loop, three--loop \cite{MSI,MSF} and
four--loop \cite{MSS} RG expansions for 3D cubic model resummed
by means of the generalized Pade--Borel (Chisholm--Borel) method
lead to the estimates:

\begin{equation}
n_c^{(2)} = 3.01, \qquad \quad
n_c^{(3)} = 2.95, \qquad \quad
n_c^{(4)} = 2.90.
\label{pp}
\end{equation}

These values being very close to their counterparts resulting
from the $\epsilon$--expansion Eq.~(\ref{o}) also go down when
the order
of the approximation grows up, and the most accurate 3D estimate
available $n_c^{(4)} = 2.90$ is appreciably smaller than 3.

Pretty good agreement between the higher--order RG estimates for
$n_c$ obtained in three and $(4- \epsilon)$ dimensions enable us
to believe that both estimates are close enough to the exact
value of $n_c$. Hence, we may conclude that $n_c < 3$ and cubic
crystals with vector order parameter (3D spins) should
demonstrate, in principle, anisotropic critical behaviour with
critical exponents differing from those of the 3D Heisenberg
model. On the other hand, since the difference between $n_c$
and 3 is rather small the cubic fixed point of the RG equations
should be located in close vicinity of the Heisenberg fixed point
at the flow diagram. As a result, critical exponents describing
the anisotropic critical behaviour should be numerically close
to the critical exponents of the Heisenberg model.

\section{SUMMARY}
\label{sec:4}

In the paper, $\sqrt\epsilon$--expansions for critical exponents
of the weakly--disordered Ising model are calculated up to the
five--loop order. Coefficients of the expansions obtained are
found to exhibit rather irregular behaviour preventing these
series from to be resummed by means of the procedures based on
the Borel transformation. This fact may be thought of as a
reflection of Borel non--summability of such expansions 
conjectured earlier on the base of studying of much simpler
(zero--dimensional) model.
The marginal spin dimensionality $n_c$ for a cubic model is
calculated by the Pade--Borel resummation of corresponding
five--loop $\epsilon$--expansion obtained by Kleinert and
Schulte--Frohlinde. The estimate $n_c = 2.855$ thus obtained
is found to agree well with its analog extracted earlier from
four--loop 3D RG expansions. The conclusion is made that the
exact value of $n_c$ is smaller than 3 and cubic crystals with
vector order parameters should demonstrate, in principle,
anisotropic critical behaviour.

\acknowledgements

We are grateful to I.~O.~Maier who has attracted our attention to
the paper \cite{HK}. One of the authors (BNS) has much
benefitted from discussions with H.~Kleinert and K.~Ziegler.
The work in Electrotechnical University was supported in part
by the Foundation for Intellectual Collaboration (St. Petersburg)
via Russian Scientific and Technological Program ``Fullerenes and
Atomic Clusters'', Project No. 94024.

\end{document}